      [1999/12/01 v1.4c Il Nuovo Cimento]
\documentclass{cimento}

\usepackage{graphicx}  
\title{Search for $\nu_{\mu}\rightarrow\nu_{\tau}$ oscillations in appearance mode in the OPERA experiment}
\author{U.~Kose\from{ins:pd}\thanks{On behalf of OPERA Collaboration, umut.kose@cern.ch}
}
\instlist{\inst{ins:pd} Dipartimento di Fisica dell Universit`a di Padova, I-35131 Padova, Italy\\
  INFN Sezione di Padova, I-35131 Padova, Italy}

\PACSes{
\PACSit{13.15.+g, 14.60.Pq, 29.40Gx, 29.40.Rg}{\ldots}}

\begin{document}

\maketitle
\begin{abstract}
The OPERA experiment in the underground Gran Sasso Laboratory (LNGS) has been designed
to perform the first detection of neutrino oscillations in direct appearance mode in the muon neutrino to tau
neutrino channel. The detector is hybrid, being made of an emulsion/lead target and of electronic
detectors. It is placed in the CNGS neutrino beam 730 $km$ away from the neutrino source. Runs
with CNGS neutrinos were successfully carried out in 2008, 2009, and 2010. After a brief description
of the beam and the experimental setup, we report on event analysis of a
sample of events corresponding to $1.89\times 10^{19}$ $p.o.t.$ in the CERN CNGS $\nu_{\mu}$ beam that yielded the
observation of a first candidate $\nu_{\tau}$ CC interaction.  The topology and kinematics of this candidate
event are described in detail. The background sources are explained and the significance of the candidate is assessed.


\end{abstract}

\section{Introduction}

Two types of experimental methods can be used to detect
neutrino oscillations: observing the appearance of a neutrino
flavour initially absent in the beam or measuring the disappearance
rate of the initial flavour. In the latter case, one must know the flux of the beam precisely. In this type of experiment
one explores whether less than the expected number of neutrinos of a produced flavour arrives at a 
detector or whether the spectral shape changes if observed at various distances from a source. 
 Since the final state is not observed, disappearance
experiments cannot tell into which flavor a neutrino has oscillated. An appearance experiment searches 
for possible new flavours of neutrino, which does not exist in the original beam, 
or for an enhancement of an existing neutrino flavour. The identification of the flavour relies 
on the detection of the corresponding lepton produced in its charged current (CC) interactions: 
$\nu_{l}N\rightarrow{l}^{-}X$
with $l= e, \mu, \tau$ and where X denotes the hadronic final state.

In the past two decades, several experiments carried out with atmospheric and accelerator neutrinos, as
well as with solar and reactor neutrinos, have established the picture
of a three-neutrino oscillation scenario with two large mixing angles. 
Atmospheric sector flavor conversion
was first established by the Super-Kamiokande~\cite{ref:sk} and MACRO~\cite{ref:macro} experiments and then confirmed 
by the K2K~\cite{ref:k2k} and MINOS~\cite{ref:minos} longbaseline
experiments. The CHOOZ~\cite{ref:chooz} and Palo Verde~\cite{ref:paloverde} reactor
experiments excluded indirectly the $\nu_{\mu}\rightarrow\nu_{e}$ channel as the
dominant process in the atmospheric sector. However, the direct
observation of flavour transition through the detection of the
corresponding lepton has never been observed. Appearance of $\nu_{\tau}$
will prove unambiguously that $\nu_{\mu}\rightarrow\nu_{\tau}$ oscillation is the dominant transition channel at the atmospheric scale.

The OPERA experiment~\cite{ref:opera} has been designed to directly observe the appearance of $\nu_{\tau}$ in a pure
$\nu_{\mu}$ beam on an event by event basis.
The $\nu_{\tau}$ signature is given by the decay topology and kinematics of the short lived $\tau^{-}$ leptons produced in the interaction 
of $\nu_{\tau}N\rightarrow\tau^{-}X$ and decaying to one prong ($\mu, e$ or $hadron$) or three prongs, which are~\cite{ref:pdg}:
\begin{eqnarray}
\tau^{-} \rightarrow \mu^{-} \nu_\mu \bar{\nu}_\tau \hspace{1.0cm} with \hspace{0.5cm} BR = 17.36 \pm {0.05}\% \hspace{4.3cm} \nonumber \\ 
\tau^{-} \rightarrow e^{-} \nu_e \bar{\nu}_\tau  \hspace{1.0cm} with \hspace{0.5cm} BR = 17.85 \pm {0.05}\% \hspace{4.4cm} \nonumber \\
\tau^{-} \rightarrow h^{-} (n\pi^{0}) \bar{\nu}_\tau \hspace{0.5cm} with \hspace{0.5cm} BR = 49.52 \pm {0.07}\% \hspace{4.3cm} \nonumber \\
\tau^{-} \rightarrow 2h^{-}h^{+} (n\pi^{0}) \bar{\nu}_\tau \hspace{0.5cm} with \hspace{0.5cm} BR = 15.19 \pm {0.08}\%.\hspace{3.6cm}  \nonumber 
\end{eqnarray}

\section{The Neutrino Beam}

The CNGS $\nu_{\mu}$ beam produced by the CERN-SPS is directed towards the
OPERA detector, located in the Gran Sasso underground laboratory (LNGS)~\cite{ref:lngs} in Italy,
730 km away from the neutrino source at CERN. In order to study $\nu_{\mu}\rightarrow{\nu}_{\tau}$ oscillations in appearance mode as indicated in the atmospheric neutrino sector, the 
CERN Neutrinos to GranSasso (CNGS) neutrino beam~\cite{ref:cngs} was designed and optimized by maximizing the number of $\nu_{\tau}$CC interactions at the LNGS.

The average $\nu_{\mu}$ beam energy
is 17 $GeV$, well above tau production energy treshold. The $\bar{\nu}_{\mu}$ contamination is $\sim 4\%$ in flux, $2.1\%$ in terms of interactions. The 
$\nu_{e}$ and $\bar{\nu}_{e}$ contaminations are lower than $1\%$, while the number of prompt $\nu_{\tau}$ from
$D_{s}$ decay is negligible. The average $L/E_{\nu}$ ratio is 43 $km/GeV$, suitable for oscillation studies at atmospheric
$\Delta{m^2}$. Due to the Earth's curvature
neutrinos from CERN enter the LNGS halls with an angle of about $3^{\circ}$ with respect to the horizontal
plane.

With a nominal CNGS
beam intensity of $4.5\times10^{19}$ protons on target ($p.o.t.$) per year, and assuming $\Delta{m}^{2}_{23}=2.5\times10^{-3} eV^{2}$ and full mixing, 
about 10 $\nu_{\tau}$ events are expected to be observed in OPERA in 5 years of
data taking, with selection criteria reducing the background to 0.75 events.


The goal is to accumulate a statistics of neutrino interactions correspomding to $22.5\times10^{19}$ $p.o.t.$ in 5 years. 
The 2008, 2009 and 2010 runs achieved a total intensity of $1.78\times10^{19}$, $3.52\times10^{19}$ and $4.04\times10^{19}$  $p.o.t.$  respectively.
Within these three years, neutrinos produced 9637 beam events. The processing of these events, particularly the scanning of emulsion films, is
continuously going on.
The 2011 run started on May 2011 and is still in progress.

At the CNGS energies the average $\tau^{-}$ decay length is submillimetric, so OPERA uses nuclear
emulsion films as high precision tracking device in order to be able to detect such short decays.
Emulsion films are interspaced with $1 mm$ thick lead plates, which act as neutrino target and form
the largest part of the detector mass. This technique is called Emulsion Cloud Chamber (ECC). It was successfully used to
establish the first evidence for charm in cosmic rays interactions~\cite{ref:niu} and in the DONUT experiment~\cite{ref:donut} for the first direct observation
of the $\nu_{\tau}$. To date, nine $\nu_{\tau}$ CC interactions have been observed by DONUT produced by a fixed
target 800 $GeV$ proton beam configuration.

\section{The OPERA Detector}

\begin{figure}[htb]
\includegraphics[width=30pc,height=20pc,scale=0.8]{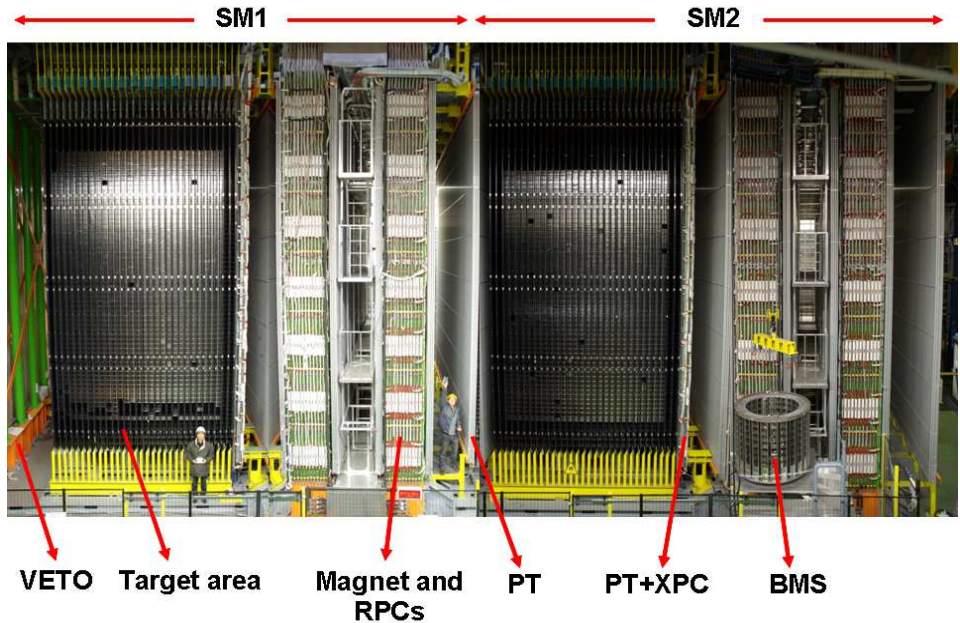}
\caption{View of the OPERA detector; the neutrino beam enters from the left. Arrows show the
position of detector components, the VETO planes, the target and TT, the drift tubes (PT) laid out along
the XPC, the magnets and the RPC installed between the magnet iron slabs. The Brick Manipulator System (BMS) is partly shown.}
\label{fig:detector}
\end{figure}

OPERA is a hybrid detector made of two identical Super Modules (SM1 and SM2), each
one formed by a target section and a muon spectrometer as shown in Figure~\ref{fig:detector}.
Each target section is organized
in 31 vertical "$walls$", transverse to the beam direction.
Walls are filled with "$ECC$ $bricks$" with an
overall mass of 1.25 $kton$. They are followed by
double layers of scintillator planes acting as Target
Trackers (TT) that are used to locate neutrino
interactions occurred within the target.
A target brick consists of 56 lead
plates of 1 mm thickness interleaved with 57
emulsion films. 
The lead plates serve as neutrino interaction
target and the emulsion films as 3-dimensional tracking
detectors providing track coordinates with a sub-micron
accuracy and track angles with a few mrad accuracy.
The material of a brick along the beam direction corresponds
to about 10 radiation length and 0.33 interaction
length. The brick size is $10cm\times12.5cm\times8cm$ and its weight is
about $8.3 kg$.

In order to
reduce the emulsion scanning load, Changeable
Sheets (CS)~\cite{ref:cs} film interfaces have been used.
They consist in tightly packed doublets of
emulsion films glued to the downstream face of
each brick. Charged particles from a neutrino
interaction in a brick cross the CS and produce
signals in the TT that allow the corresponding
brick to be identified and extracted by an
automated Brick Manipulator System (BMS). 

\begin{figure}[htb]
\includegraphics[width=16pc,height=12pc,scale=0.8]{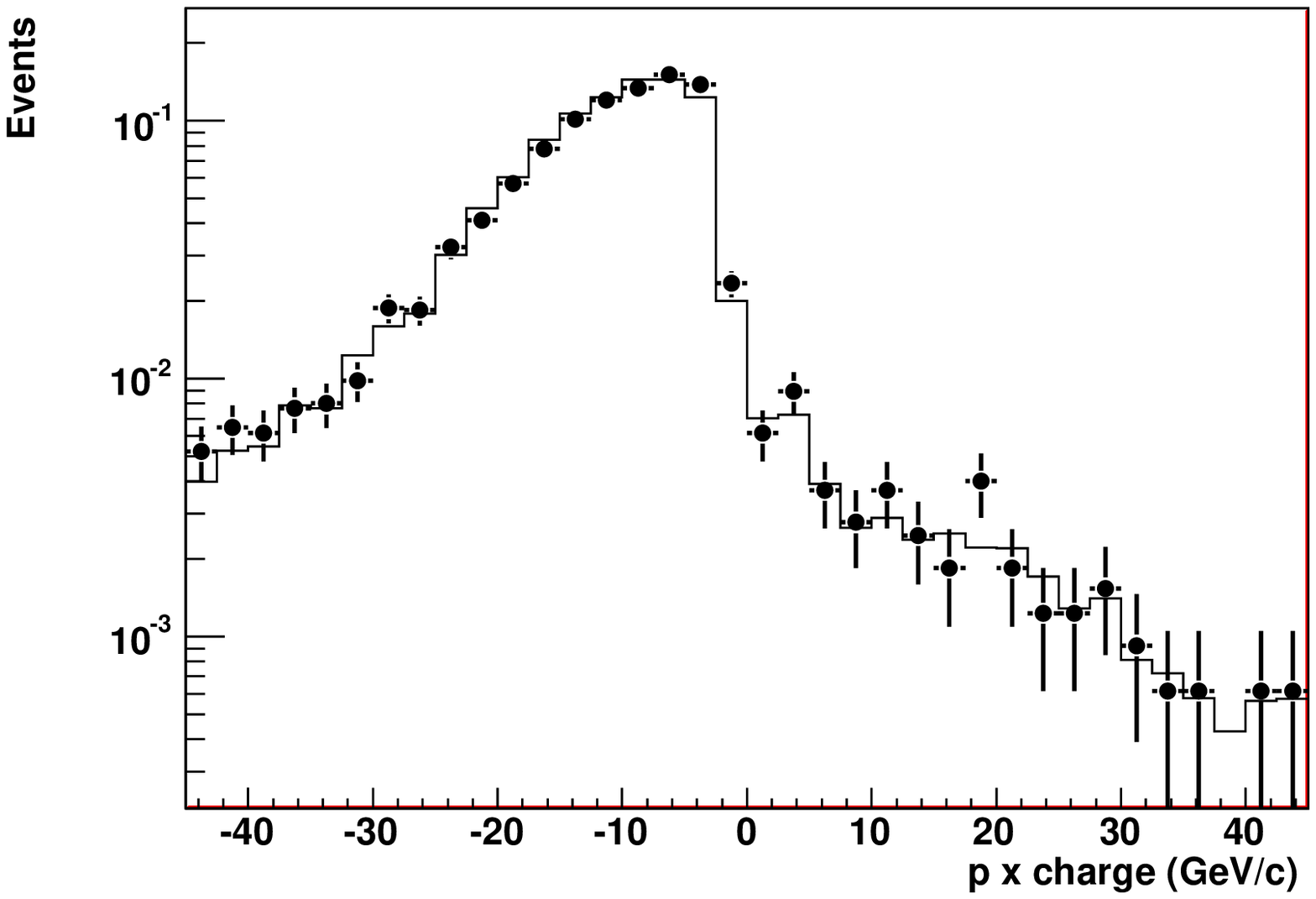}
\includegraphics[width=16pc,height=12pc,scale=0.8]{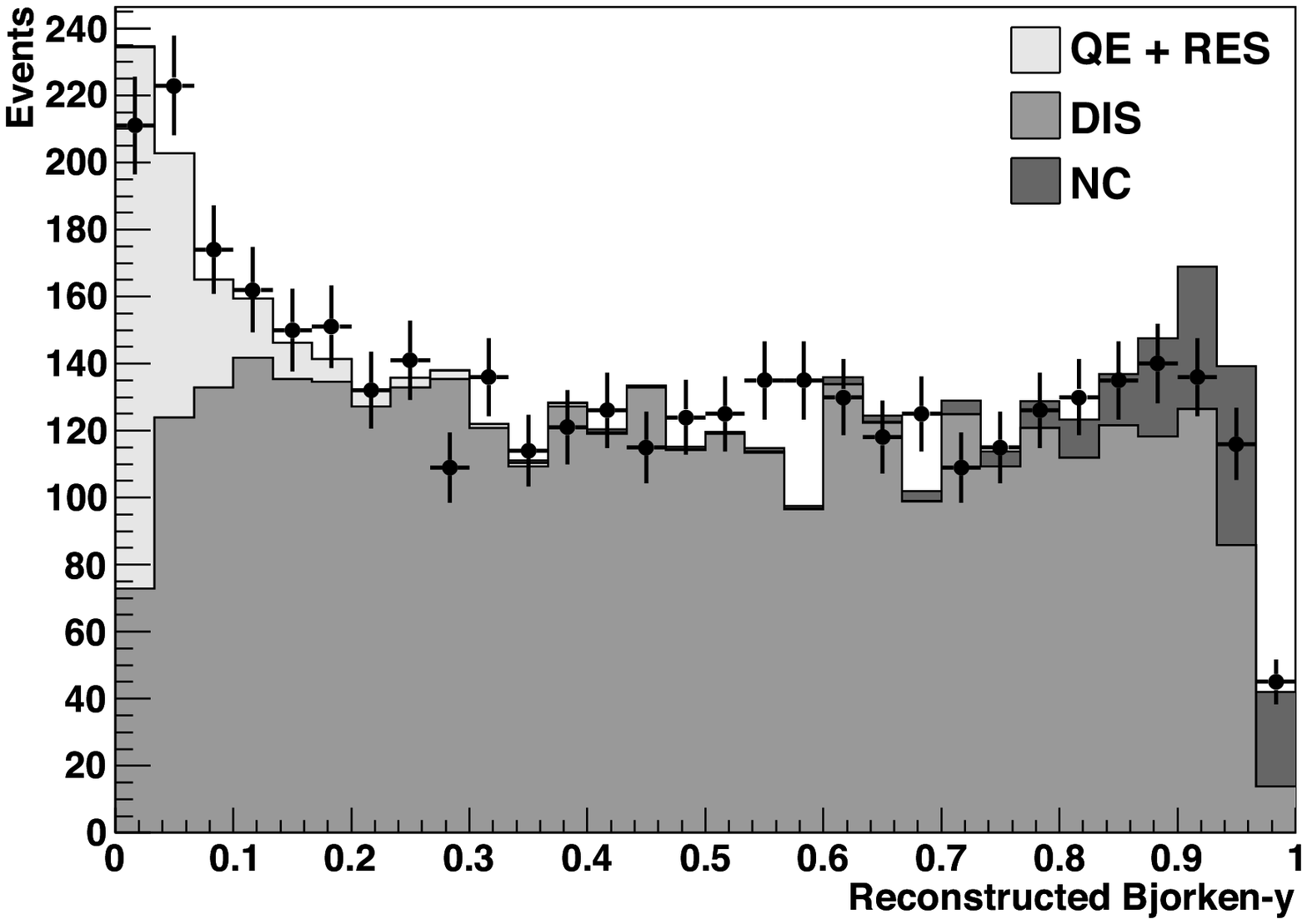}
\caption{Right: Muon charge comparison (momentum$\times$charge): data (black dots with error bars) and MC
(solid line) are normalised to one. Left: Bjorken-y variable reconstructed in data (dots with error bars) and MC (shaded areas). The
MC distributions are normalised to data. The different contributions of the MC are shown in different
colours: QE + RES contribution in light grey, DIS contribution in grey and the NC contamination in
dark grey.}
\label{fig:mom}
\end{figure}

The spectrometers consist of a dipolar magnet instrumented
with active detectors, planes of RPCs
(Internal Tracker, IT) and drift tubes (Precision
Tracker, PT). Tasks of the spectrometers are
muon identification and charge measurement in
order to minimize the background. For muon momenta between 2.5 $GeV/c$ and 45 $GeV/c$, the fraction of events with
wrong charge determination is $1.2\%$. The $\mu^{+}$ to $\mu^{-}$ events ratio, within the selected momentum range,
obtained from data can be directly compared with predictions based on Monte Carlo simulations: $3.92\pm0.37(stat.)\%$ for data, $3.63\pm0.13 (stat.) \%$ for MC. 
Figure~\ref{fig:mom} left-side shows the momentum and momentum times charge distribution for data and MC.

In Figure~\ref{fig:mom} right-side, Bjorken-y distribution is shown for the events with at
least a muon track. The agreement between data and MC simulation is reasonable. The sum of the QE and RES
processes can be clearly seen as a peak at low y values. The NC contribution shows up at values of
Bjorken-y close to one. The NC contribution becomes negligible when a track with its momentum measured by the spectrometer is required. 

A detailed description of the complete detector can be found in~\cite{ref:opera}. Event reconstruction procedures
 and a performances of the OPERA electronic detectors can be found in more detail in~\cite{ref:detector}.

\section{Neutrino interaction location}

Neutrino event analysis starts with the pattern recognition in the electronic detectors. Charged particle tracks
produced in a neutrino interaction generate signals in the TT and in the muon spectrometer. 
A brick finding algorithm is applied in order to select the brick which has the
maximum probability to contain the neutrino interaction. The
brick with the highest probability is extracted from the detector for analysis. The efficiency of this procedure reaches
$83\%$ in a subsample where up to 4 bricks per event were processed.

After extraction of the brick predicted by the
electronic detectors, its validation comes from the analysis of the CS films. 
The measurement of emulsion films is performed
through high-speed automated microscopes~\cite{ref:ess,ref:suts} with a sub-micrometric position
resolution and angular resolution of the order of
one milliradian. 
If no expected charged track related to the event is found in the CS, the brick is returned back to
the detector with another CS doublet attached.
If any track originating from the interaction is
detected in the CS, the brick is exposed to cosmic
rays (for alignment purposes) and then depacked.
The emulsion films are developed and
sent to the scanning laboratories of the Collaboration
for event location studies and decay search
analysis. 

All the track information of the CS is then used for a precise
prediction of the tracks in the most
downstream films of the brick (with an accuracy of about $100\mu{m}$). When found in this films, tracks are followed upstream
from film to film. The scan-back procedure
is stopped when no track candidate is found in
three consecutive films and the lead plate just upstream
the last detected track segment is defined as the vertex
plate. In order to study the located vertices and reconstruct the events, 
a general scanning volume is defined with a transverse area of $1\times1cm^{2}$ for 5 films upstream and 10 films
downstream of the stopping point. All track segments in this volume are collected and analysed. After rejection of
the passing through tracks related to cosmic rays and of the tracks due to low energy particles, the tracks produced by the neutrino
interaction can be selected and reconstructed.

The present overall
location efficiency averaged over NC and CC
events, from the electronic detector predictions
down to the vertex confirmation, is about $60\%$.

\section{Decay Search}

Once the neutrino interaction is located, a decay search procedure is applied to detect
possible decay or interaction topologies on tracks attached to the primary vertex.
The main signature of a secondary vertex (decay or
nuclear inetaraction) is the observation of a track with
a significant impact parameter (IP) relative to the neutrino
interaction vertex. The IP of primary tracks is smaller than 10$\mu{m}$ after excluding tracks produced by low momentum particles.
When secondary vertices
are found in the event, a kinematical analysis
is performed, using particle angles and momenta
measured in the emulsion films. For charged particles
up to about 6 $GeV/c$, momenta can be determined
using the angular deviations produced
by Multiple Coulomb Scattering (MCS) of tracks
in the lead plates~\cite{ref:mcs} with a resolution better
than $22\%$. 
For higher momentum particles, the measurement is based
on the position deviations. The resolution is better
than $33\%$ on 1/p up to 12 $GeV/c$ for particles
passing through an entire brick.

A $\gamma$-ray search is performed in
the whole scanned volume by checking all tracks
having an IP with respect to
the primary or secondary vertices lower than 800$\mu{m}$. The angular acceptance is $\pm500$ $mrad$. The
$\gamma$-ray energy is estimated by a Neural Network algorithm
that uses the number
of segments, the shape of the electromagnetic
shower and also the MCS of the leading tracks.

\section{Data analysis}

In the following, the analysis results~\cite{ref:taupaper} of about $35\%$ of the 2008 and 2009 data sample, corresponding 
to the $1.89\times10^{19}$ $p.o.t$ are presented. The decay search procedure was
applied to a sample of 1088 events of which 901 were classified as CC interactions. In the sample of CC
interactions, 20 charm decay candidates were
observed, in good agreement with the expectations from the Monte Carlo
simulation, $16\pm 2.9$. Out of them 3 have a 1-prong topology where $0.8\pm0.2$ was expected. The background for the total charm
sample is about 2 events. 
Several $\nu_{e}$-induced events have also been observed.

Moreover, a first CC $\nu_{\tau}$ candidate has been detected. The expected number of $\nu_{\tau}$ events detected in the
analysed sample is about $0.54\pm0.13(syst.)$ at $\Delta{m^{2}}_{23}=2.5\times10^{-3}$ $eV^{2}$ and full mixing.

\section{The first tau neutrino candidate}

\begin{figure}[htb]
\includegraphics[width=16pc,height=12pc,scale=0.8]{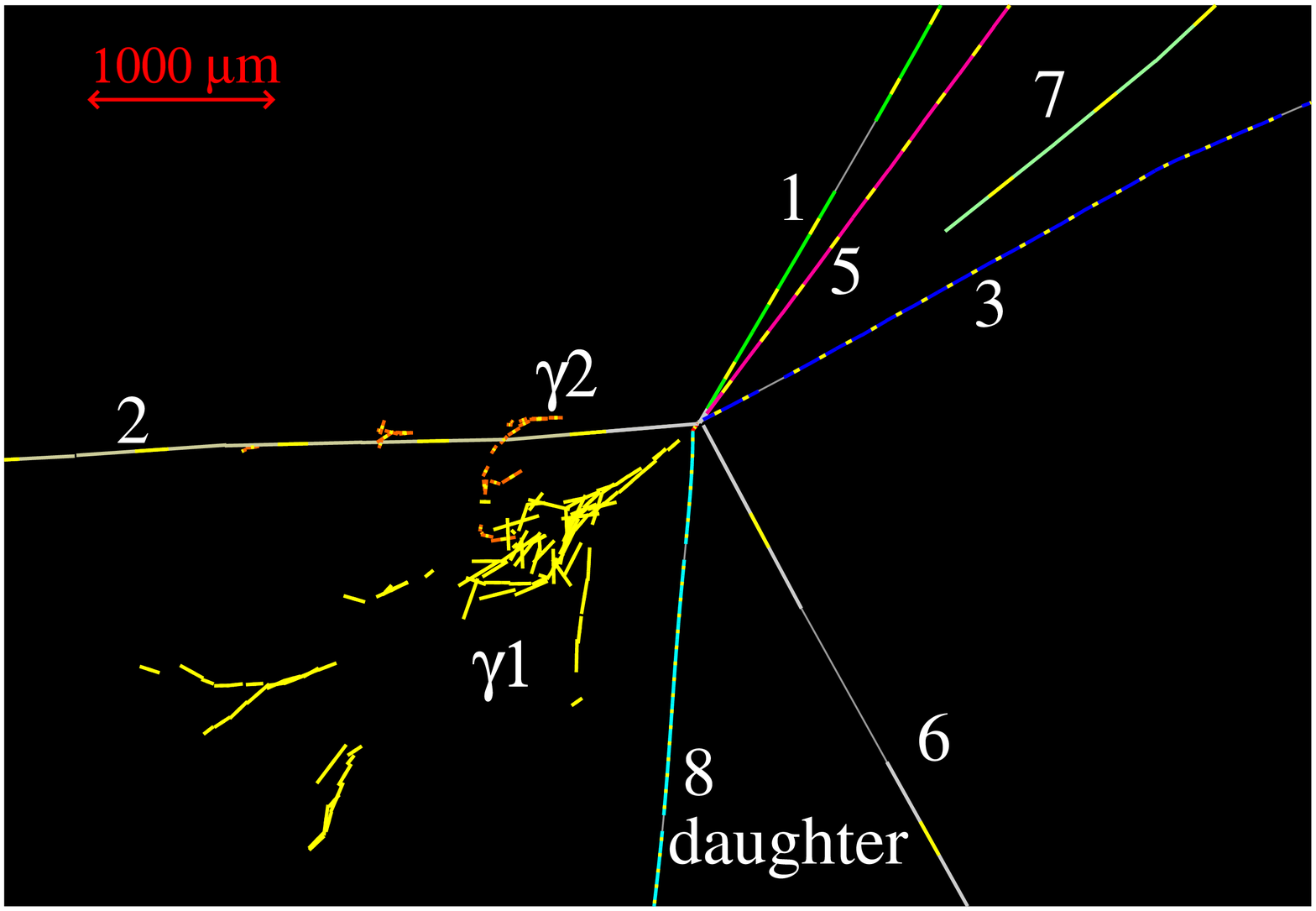}
\includegraphics[width=16pc,height=12pc,scale=0.8]{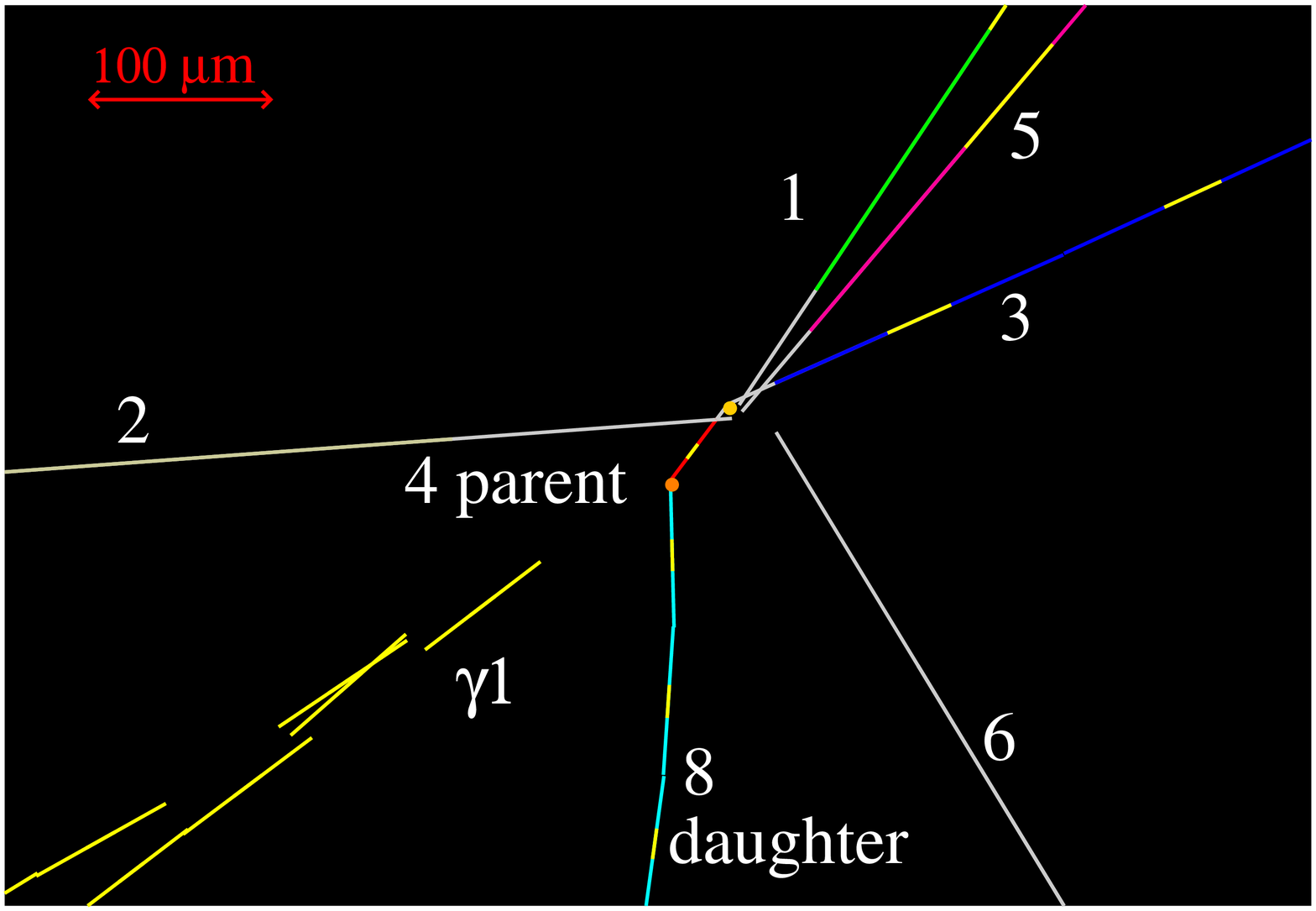}
\caption{Display of the $\nu_{\tau}$ candidate event. Left: view
transverse to the neutrino direction. Right: same view zoomed on the
vertices.The short track named "4 parent" is the $\tau^{-}$
candidate.}
\label{fig:tau}
\includegraphics[width=32pc,height=15pc,scale=0.8]{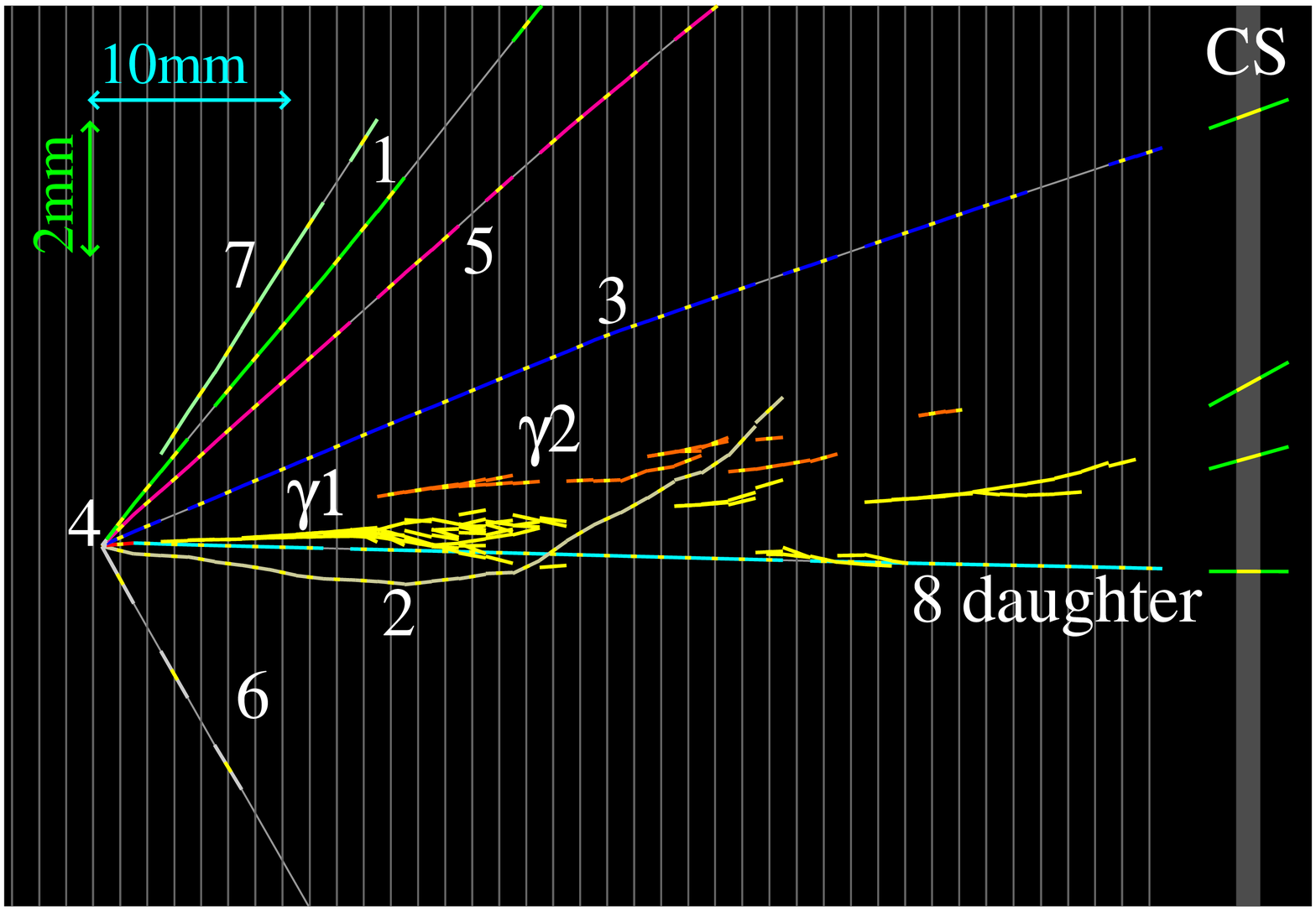}
\caption{Longitudinal view of the $\nu_{\tau}$ candidate event.}
\label{fig:tau2}
\end{figure}

In this section, the first tau neutrino candidate~\cite{ref:taupaper} will
be described. The location and decay search procedure yielded a neutrino interaction vertex with 7 tracks.
One track exhibits a visible kink with an angular change of $41\pm2 mrad$ after a path length of $1335\pm35 \mu{m}$. The kink daughter momentum is estimated to be 
$12^{+6}_{-3}$ $GeV/c$ by MCS measurement and its transverse momentum to the parent direction is $470^{+230}_{-120}$ $MeV/c$.
The event is displayed in Figures~\ref{fig:tau} and ~\ref{fig:tau2}.

All the tracks from the neutrino interaction vertex were followed until they stop or interact. The probability that one of them is left
by a muon is estimated to be less than $10^{-3}$. The residual probability for being a $\nu_{\mu}$CC event, with a possibly undetected
large angle $\mu$ track, is about $1\%$; a nominal value of $5\%$ is assumed. None of the tracks is compatible with being an electron.

Two electromagnetic showers caused by $\gamma$-rays, associated with the event, have been located and studied.
The energy of $\gamma1$ is $(5.6\pm1.0(stat.)\pm1.7(syst.))$ $GeV$ and it is clearly pointing to the decay vertex.
The $\gamma2$ has an energy of $1.2\pm0.4(stat.)\pm0.4(syst.)$ $GeV$ and it is compatible with pointing to either vertex, 
with a significantly larger probability to the decay vertex. 

All the selection cuts used in the analysis were those described in detail in the experiment
proposal~\cite{ref:proposal} and its addendum~\cite{ref:addendum}. All the kinematical variables of the event and the cut applied are given in
Table~\ref{tab:kinematic}. 

\begin{table}
  \caption{Kinematical variables of $\nu_{\tau}$ candidate event.}
  \label{tab:kinematic}
  \begin{tabular}{lcl}
    \hline
Variable                &   Measured          & Selection criteria  \\
    \hline
Kink angle ($mrad$)     & $42\pm 2$           &  \textgreater 20           \\
Decay length ($\mu{m}$) & $1335\pm 35$        & Within 2 plates     \\
P daughter ($GeV/c$)    & $12^{+6}_{-3}$      & \textgreater 2             \\
PT daughter ($MeV/c$)   & $470^{+230}_{-120}$ & \textgreater 300 ($\gamma$ attached) \\
Missing PT ($MeV/c$)    & $570^{+320}_{-170}$ & \textless 1000           \\
Angle $\phi$ ($deg$)    & $173\pm 2$          & \textgreater 90             \\
    \hline
  \end{tabular}
\end{table}

The invariant mass of the two observed $\gamma$-rays
is $120 \pm 20 (stat) \pm 35 (syst)$ supporting the hypothesis that they are
emitted in a $\pi^{0}$ decay. The invariant mass of the charged
decay daughter assumed to be a $\pi^{-}$
and of the two $\gamma$-rays amount to $640^{+125}_{-80} (stat)^{+100}_{-90} (syst)$ $MeV/c$, which is compatible with
the $\rho(770)$ mass. So the decay mode of the candidate is consistent with the hypothesis
$\tau^{-}\rightarrow\rho^{-}\nu_{\tau}$ (where the branching ratio is about $25\%$).

\section{Background Estimation}
The two main sources of
background to the $\tau^{-}\rightarrow{h}(n\pi^{0})\nu_{\tau}$ channel
where a similar final state may be produced are:
\begin{itemize}
\item the decays of charmed particles produced in $\nu_{\mu}$ CC
interactions where the primary muon is not
identified as well as the $c\bar{c}$ pair production
in $\nu_{\mu}$ NC interactions where one charm particle
is not identified and the other decays to a 1-prong
hadron channel; 

\item the 1-prong
inelastic interactions of primary hadrons
produced in $\nu_{\mu}$CC interactions where the
primary muon is not identified or in $\nu_{\mu}$ NC
interactions and in which no nuclear fragment
can be associated with the secondary
interaction. 
\end{itemize}
The Monte Carlo expectation of the first background source is $0.007\pm0.004(syst.)$ event, the fraction produced in $\nu_e$ CC
interactions is less than $10^{-3}$ events, The second type of
background amounts to $0.011\pm0.006(syst.)$ event. The total background in the decay channel to a single 
charged hadron is $0.018 \pm 0.007 (syst)$ events.
The probability that this background events
fluctuate to one event is $1.8\%$ $(2.36 \sigma)$. As the
search for $\tau^{-}$ decays is extended to all four
channels, the total background then becomes
$0.045 \pm 0.023 (syst)$. The probability that this
expected background to all searched decay
channels of the $\tau^{-}$ fluctuates to one event is $4.5\%$ $(2.01 \sigma)$. At $\Delta{m}^{2} = 2.5 \times 10^{-3}$ $eV^{2}$ and full
mixing, the expected number of observed $\tau^{-}$
events with the present analyzed statistics is
$0.54 \pm 0.13 (syst)$ of which $0.16 \pm 0.04 (syst)$ in
the one-prong hadron topology, compatible
with the observation of one event.

\section{Conclusions}

During 2008, 2009 and 2010 runs, a total intensity of $1.78\times10^{19}$, $3.52\times10^{19}$ and 
$4.04\times10^{19}$  $p.o.t.$ respectively, was achieved. Within these three years, 9637 beam events have been collected within the OPERA target. 
The neutrino interaction location and decay search are going on.

A first candidate $\nu_{\tau}$ CC interaction
in the OPERA detector at LNGS was detected
after analysis of a sample of events corresponding
to $1.89\times10^{19}$ $p.o.t.$ in the CERN CNGS
$\nu_{\mu}$ beam. The expected number of $\nu_{\tau}$ events in
the analysed sample is $0.54\pm0.13 (syst.)$. The candidate
event passes all selection criteria, it is
assumed to be a $\tau^{-}$ lepton decaying into $h^{-}(n\pi^{0})\nu_{\tau}$.
The observation of one possible tau candidate in
the decay channel $h^{-}(\pi^{0})\nu_{\tau}$ has a significance of
$2.36\sigma$ of not being a background fluctuation.

\end{document}